\documentclass [12 pt]{article}
\def\be{\begin{equation}}
\def\eea{\end{eqnarray}}
\def\bea{\begin{eqnarray}}
\def\ee{\end{equation}}

\usepackage[psamsfonts]{amssymb}
\usepackage{amsmath}
\author{M. Amooshahi$^{1,2}$ \footnote{amooshahi@sci.ui.ac.ir} and F.
Kheirandish $^{2}$ \footnote{fardin$_{-}$kh@phys.ui.ac.ir}
\\ $^{1}$ {\small Faculty of science, Shahrekord University, Shahrekord, Iran}
\\$^{2}$ {\small Faculty of science, University of Isfahan, Hezar Jarib Ave., Isfahan,Iran}}

\title{Electromagnetic field quantization in an anisotropic and inhomogeneous
 magnetodielectric}
\begin{document}
\maketitle
\begin{abstract}

The electromagnetic field in an anisotropic and inhomogeneous
magnetodielectric is quantized by modelling the medium with two
independent quantum fields. Some coupling tensors coupling the
electromagnetic field with the medium are introduced. Electric and
magnetic polarizations are obtained in terms of the ladder
operators of the medium and the coupling tensors explicitly. Using
a minimal coupling scheme for electric and magnetic interactions,
the Maxwell equations and the constitutive equations of the medium
are obtained. The electric and magnetic susceptibility tensors of
the medium are calculated in terms of the coupling tensors.
Finally the efficiency of the approach is elucidated by some
examples.



\end{abstract}
\section{Introduction}
The quantization of electromagnetic field in an  absorptive
dielectric, represents one of the most and interesting problems in
quantum optics, because it gives a rigorous test of our
understanding of the interaction of light with matter. One of the
important methods to quantize the electromagnetic field in the
presence of an absorptive medium is known as Green function method
\cite{.1}-\cite{.7}. In this method by adding the noise electric
and magnetic polarization densities to classical constitutive
equations of the medium, these equations are considered as
definitions of electric and magnetic polarization operators. The
noise polarizations are related to two independent sets of bosonic
operators. Combination of the Maxwell equations and the
constitutive equations in frequency domain, give the
electromagnetic field operators in terms of the noise
polarizations and classical Green tensor. Suitable commutation
relations are imposed on the bosonic operators such that the
commutation  relations between electromagnetic field operators
become identical with those in
free space.\\
An interesting quantization scheme of electromagnetic field in the
presence of an absorptive dielectric medium is based on the
Hopfield model of a dielectric \cite{1}, where the polarization of
the dielectric is represented by a damped quantum field \cite{2}.
Huttner and Barnett \cite{3} for a homogeneous medium and after
Suttorp and Wubs \cite{4} for an inhomogeneous medium in the
framework of the damped polarization model have presented a
canonical quantization for the electromagnetic field inside an
absorptive dielectric. This scheme is based on a microscopic model
in which the medium is represented by a collection of interacting
matter fields. The absorptive character of the medium is modelled
through the interaction of the matter fields with a reservoir
consisting of a continuum of the Klein-Gordon fields. In this
model, eigen-operators for the coupled systems are calculated and
the electromagnetic field operators have been expressed in terms
of these eigen-operators. Also, the dielectric function is derived
and it is shown that it satisfies
 the Kramers-Kronig relations \cite{5}.\\
Another approach to quantizing a dissipative system is by
considering the dissipation as a result of interaction between the
system and a heat bath consisting of a set of harmonic oscillators
\cite{5.9}-\cite{17}. In this method the whole system is composed
of two parts, the main system and a heat bath which interacts with
the main system and causes the dissipation of energy on it.

 In a recent approach to electromagnetic field quantization the present authors have quantized the electromagnetic
  field in an isotropic magnetodielectric \cite{18}. In this approach: (i) the electromagnetic field is taken as the main
  quantum system and the medium as a heat bath. (ii) The polarizability of the medium is defined in terms of
   dynamical variables of the medium. (iii) The polarizability and absorptivity of the
medium are not independent of each other, as expected, this is
contrary to the damped polarization model where polarizability and
absorptivity are treated independently \cite{3,4}. (iv) If the
medium is both magnetizable and polarizable, one must models the
medium with two independent collections of harmonic oscillators,
where one collection describes electric properties and the other
one describes magnetic properties of the medium. This scheme leads
to a consistent quantization of the electromagnetic field in the
presence of an absorptive magnetodielectric \cite{18}.\\
In the present article, the idea introduced in the previous work
\cite{18} is generalized to the case of an anisotropic and
inhomogeneous magnetodielectric.

\section{ Quantum dynamics}

Electromagnetic field quantization can be achieved in an
anisotropic magnetodielectric by modelling the medium with two
independent quantum fields. Let us call these fields E and M
quantum fields, describing the polarizability and magnetizability
of the medium respectively. These quantum fields couple the medium
with electromagnetic field through some coupling tensors. The
electric and magnetic polarization densities of the medium are
defined as linear expansions in terms of the ladder operators of
the E and M fields. The coefficients of these expansions are real
valued coupling tensors. We will see that the electric and
magnetic susceptibility tensors can be obtained in terms of the
coupling tensors. In the following we use the Coloumb gauge and
assume the periodic boundary conditions with no loss of generality
of the approach.

The electromagnetic vector potential $\vec{A}$ inside a box with
volume $V=L_1L_2L_3$ can be expanded in terms of plane waves as
\begin{equation} \label{d1}
\vec{A}(\vec{r},t)=\sum_{\vec{n}} \sum_{\lambda=1}^2
\sqrt{\frac{\hbar}
{2\varepsilon_0V\omega_{\vec{n}}}}\left[a_{\vec{n}\lambda}(t)e^{i\vec{k}_{\vec{n}}\cdot\vec{r}}+a_{\vec{n}
\lambda}^\dag(t)e^{-i\vec{k}_{\vec{n}}\cdot\vec{r}}\right]\vec{e}_{\vec{n}\lambda},
\end{equation}
where $\omega_{\vec{n}}=c|\vec{k}_{\vec{n}}|$ is the frequency
corresponding to the mode $\vec{n}$, the vector $\vec{n}$ is a
triplet of integer numbers $ (n_1,n_2,n_3) $, $ \sum_{\vec{n}}$
means $\displaystyle\sum_{n_1,n_2,n_3=-\infty}^{+\infty} $,
$\varepsilon_0$ is the permittivity of the vacuum,
$\vec{k}_{\vec{n}}=\frac{2n_1\pi}{L_1}\vec{i}+\frac{2n_2\pi}{L_2}\vec{j}+\frac{2n_3\pi}{L_3}\vec{k}$
 is the wave  vector and
$\vec{e}_{\vec{n}\lambda}$ are polarization unit vectors for each
$\vec{n}$, satisfying

\begin{eqnarray}\label{d2}
\vec{e}_{\vec{n}\lambda}.\vec{e}_{\vec{n}\lambda'}&=&\delta_{\lambda\lambda'},\nonumber\\
 \vec{e}_{\vec{n}\lambda}.\vec{k}_{\vec{n}}&=&0.
\end{eqnarray}

The operators $a_{\vec{n}\lambda}(t)$ and
$a_{\vec{n}\lambda}^\dag(t) $ are annihilation and creation
operators of the electromagnetic field and satisfy the following
equal time commutation rules
\begin{equation}\label{d3}
[a_{\vec{n}\lambda}(t),a_{\vec{m}\lambda'}^\dag(t)]=
\delta_{\vec{n},\vec{m}}\delta_{\lambda\lambda'}.
\end{equation}
Quantization in Coloumb gauge usually needs resolution of a vector
field in its transverse and longitudinal parts. Any vector field
$\vec{F}(\vec{r})$ can be resolved in two components, transverse
and longitudinal component which are denoted by $ \vec{F}^\bot$
and $\vec{F}^\|$ respectively. The transverse part satisfy the
coloumb condition $ \nabla\cdot\vec{F}^\bot=0$ and the
longitudinal component is a conservative field $
\nabla\times\vec{F}^\|=0$. For a periodic boundary condition these
two parts are defined as
\begin{eqnarray}\label{d4}
&& \vec{F}^\bot(\vec{r},t)=\vec{F}(\vec{r},t)+\int_V d^3r'
\nabla'\cdot\vec{F}(\vec{r'},t)\vec{\nabla} G(\vec{r},\vec{r'}), \\
&&\vec{F}^\|(\vec{r},t)=-\int_V d^3r'
\nabla'\cdot\vec{F}(\vec{r'},t)\vec{\nabla} G(\vec{r},\vec{r'}),
\end{eqnarray}
where
\begin{equation}\label{d5}
 G(\vec{r},\vec{r'})=\sum_{\vec{n}}\frac{1}{|\vec{k}_{\vec{n}}|^2}e^{
 \imath\vec{k}_{\vec{n}}\cdot(\vec{r}-\vec{r'})},
\end{equation}

is the Green function and satisfies the Poison equation
\begin{equation}\label{d6}
\nabla^2G(\vec{r},\vec{r'})=-\delta(\vec{r}-\vec{r'}).
\end{equation}

 In absence of external charges the displacement field is  purely transverse,
and we can expand  it in terms of the plane waves
\begin{equation}\label{d7}
\vec{D}(\vec{r},t)=-i\varepsilon_0\sum_{\vec{n}}\sum_{\lambda=1}^2
\sqrt{\frac{\hbar\omega_{\vec{n}}}{2\varepsilon_0V}}
\left[a_{\vec{n}\lambda}^\dag(t)e^{-\imath\vec{k}_{\vec{n}}\cdot\vec{r}}-a_{\vec{n}\lambda}(t)e^{\imath\vec{k}_{\vec{n}}\cdot\vec{r}}\right]\vec{e}_{\vec{n}\lambda}.
\end{equation}

 The commutation relations (\ref{d3}) lead to the following commutation
relations between the components of the vector potential $\vec{A}$
and the displacement operator $\vec{D}$
\begin{equation}\label{d8}
[A_l(\vec{r},t),-D_j(\vec{r'},t)]=
\imath\hbar\delta_{lj}^\bot(\vec{r}-\vec{r'}),
\end{equation}
where
\begin{equation}\label{d9}
 \delta_{lj}^\bot(\vec{r}-\vec{r'})=\frac{1}{V}
\sum_{\vec{n}}(\delta_{lj}-\frac{k_{\vec{n}l} k_{\vec{n}j}
}{|\vec{k}_{\vec{n}}|^2})e^{\imath\vec{k}_{\vec{n}}\cdot(\vec{r}-\vec{r'})},
\end{equation}
is the transverse delta function. From (\ref{d9}), we see that
$-\vec{D}$ plays the role of the momentum density of
electromagnetic field. The Hamiltonian of the electromagnetic
field inside the box is given by

\begin{eqnarray}\label{d10}
&&H_F(t)=\int_V d^3r \left[\frac{ \vec{D}^2}{2\varepsilon_0}+
\frac{(\nabla\times\vec{A})^2}{2\mu_0}\right]=
\sum_{\vec{n}}\sum_{\lambda=1}^2\hbar
\omega_{\vec{n}}a_{\vec{n}\lambda}^\dag(t)
a_{\vec{n}\lambda}(t).\nonumber\\
&&
\end{eqnarray}

where $\mu_0$ is the magnetic permittivity of the vacuum and we
have used normal ordering for $ a_{\vec{n}\lambda}^\dag(t) $ and $
a_{\vec{n}\lambda}(t) $.

Now we include the medium in the process of quantization. For this
purpose let the Hamiltonian corresponding to E and M quantum
fields be denoted by $H_e$ and $H_m$ respectively. Then the medium
Hamiltonian can be written as

\begin{eqnarray}\label{d11}
&&H_d=H_e+H_m,\nonumber\\
&& H_e(t)=\sum_{\vec{n}}\sum_{\nu=1}^3\int_{-\infty}^{+\infty}d^3q
\hbar\omega_{\vec{q}}
d_{\vec{n}\nu}^\dag(\vec{q},t)d_{\vec{n}\nu}(\vec{q},t),
\nonumber\\
&&H_m(t)=\sum_{\vec{n}}\sum_{\nu=1}^3\int_{-\infty}^{+\infty}d^3q
\hbar\omega_{\vec{q}}
b_{\vec{n}\nu}^\dag(\vec{q},t)b_{\vec{n}\nu}(\vec{q},t),
\end{eqnarray}

where the annihilation and creation operators $
d_{\vec{n}\nu}(\vec{q},t)$, $d_{\vec{n}\nu}^\dag(\vec{q},t)$, $
b_{\vec{n}\nu}(\vec{q},t)$ and $b_{\vec{n}\nu}^\dag(\vec{q},t)$
satisfy the following equal-time commutation relations
\begin{eqnarray}\label{d12}
&&[d_{\vec{n}\nu}(\vec{q},t),d_{\vec{m}\nu'}^\dag(\vec{q}',t)]=
\delta_{\vec{n},\vec{m}}\delta_{\nu\nu'}\delta(\vec{q}-\vec{q'}),\nonumber\\
&&[b_{\vec{n}\nu}(\vec{q},t),b_{\vec{m}\nu'}^\dag(\vec{q'},t)]=
\delta_{\vec{n},\vec{m}}\delta_{\nu\nu'}\delta(\vec{q}-\vec{q'}).
\end{eqnarray}
In relations (\ref{d11}) $\omega_{\vec{q}}$ is the dispersion
relation of the magnetodielectric. It is remarkable to note that,
although the medium is anisotropic in its electric and magnetic
properties, we do not need to take the dispersion relation as a
tensor. As discussed in \cite{18}, we can assume a linear
dispersion relation $\omega_{\vec{q}}=c|\vec{q}|$ with no loss of
generality, but taking a linear dispersion relation simplifies the
formulas considerably. Therefore from now on we choose the
dispersion relation as $\omega_{\vec{q}}=c|\vec{q}|$ where
$c=\frac{1}{\sqrt{\varepsilon_0\mu_0}}$ is the proportionality
constant.

 The basic idea in this quantization method is that the
electric and magnetic properties of an anisotropic
magnetodielectric can be described by E and M quantum fields. This
means that we can define the electric and magnetic polarization
densities of a linear but anisotropic medium as linear
combinations of the ladder operators of the E and M quantum
fields, respectively. Therefore

\begin{eqnarray}\label{d13}
&&P_i(\vec{r},t)=\frac{1}{\sqrt{V}}\sum_{\vec{n}}\sum_{\nu=1}^3\int
d^3\vec{q}f_{ij}(\omega_{\vec{q}},\vec{r})\left[d_{\vec{n}\nu}(\vec{q},t)e^{\imath
\vec{k}_{\vec{n}}\cdot\vec{r}} +h.c.\right]
v^j_{\vec{n}\nu},\nonumber\\
&&
\end{eqnarray}
\begin{eqnarray}\label{d14}
&&M_i(\vec{r},t)=\frac{\imath}{\sqrt{V}}\sum_{\vec{n}}\sum_{\nu=1}^3\int
d^3\vec{q}g_{ij}(\omega_{\vec{q}},\vec{r})\left[b_{\vec{n}\nu}(\vec{q},t)e^{\imath
\vec{k}_{\vec{n}}\cdot\vec{r}} -h.c.\right]
s^j_{\vec{n}\nu},\nonumber\\
&&
\end{eqnarray}

where $\vec{P}$ and $\vec{M}$ are electric and magnetic
polarization densities of the medium and

\begin{eqnarray}\label{d15}
&&\vec{v}_{\vec{n}\nu}=\vec{e}_{\vec{n}\nu},\hspace{1.50cm}\nu=1,2\nonumber\\
&&\vec{s}_{\vec{n}\nu}=\hat{k}_{\vec{n}}\times\vec{e}_{\vec{n}\nu},\hspace{1cm}\nu=1,2\nonumber\\
&&\vec{v}_{\vec{n}3}=\vec{s}_{\vec{n}3}=\hat{k}_{\vec{n}}\hspace{1.50cm}
\hat{k}_{\vec{n}}=\frac{\vec{k}_{\vec{n}}}{|\vec{k}_{\vec{n}}|}.\nonumber\\
\end{eqnarray}

In definitions of polarization densities (\ref{d13}) and
(\ref{d14}), the real valued tensors
 $ f_{ij}(\omega_{\vec{q}},\vec{r})$ and $g_{ij}(\omega_{\vec{q}},\vec{r}) $, are called the coupling
 tensors  of the electromagnetic field and the medium which are dependent (independent) on position $ \vec{r} $ for
 inhomogeneous (homogeneous) magnetodielectrics. The coupling tensors play the key role in this method
  and are a measure for the strength of the polarizability and magnetizability of the
  medium macroscopically. We will see that the imaginary parts of the electric and magnetic
  susceptibilty in frequency domain can be obtained in terms of
  these coupling tensors. Also, explicit forms for the noise polarization densities
  can be obtained in terms of the coupling tensors and the ladder operators of the
medium. The coupling tensors are common factors in the noise
densities and the electric and magnetic susceptibilities, and so
the strength of the noise densities are dependent on the strength
of the electric and magnetic susceptibility. It can be shown that
for a non absorptive medium, the noise densities tend to zero as
expected and this quantization scheme reduces to the usual
quantization in such media.

 A consistent quantization scheme must lead to the correct equations of motion
 of the system and the medium. These equations are macroscopic Maxwell and constitutive equations of the
 medium and we will see that these equations can be obtained from the Heisenberg equations
using the total Hamiltonian defined by

\begin{eqnarray}\label{d16}
&&\tilde{H}(t)=\int d^3r \left\{\frac{[
\vec{D}(\vec{r},t)-\vec{P}(\vec{r},t)]^2}{2\varepsilon_0}+
\frac{(\nabla\times\vec{A})^2(\vec{r},t)}{2\mu_0}
-\nabla\times\vec{A}(\vec{r},t).\vec{M}(\vec{r},t)\right\}\nonumber\\
&&+H_e+H_m.
\end{eqnarray}

\subsection{Maxwell equations}
Using the commutation relations (\ref{d8}) the Heisenberg equations
for the vector potential $\vec{A}$ and the displacement field
$\vec{D}$ are
\begin{equation}\label{d17}
\frac{\partial\vec{A}(\vec{r},t)}{\partial
t}=\frac{\imath}{\hbar}[\tilde{H},\vec{A}(\vec{r},t)]=
-\frac{\vec{D}(\vec{r},t)-\vec{P}^\bot(\vec{r},t)}{\varepsilon_0},
\end{equation}
\begin{equation}\label{d18}
\frac{\partial\vec{D}(\vec{r},t)}{\partial
t}=\frac{\imath}{\hbar}[\tilde{H},\vec{D}(\vec{r},t)]=
\frac{\nabla\times\nabla\times\vec{A}(\vec{r},t)}{\mu_0}-\nabla\times\vec{M}(\vec{r},t),
\end{equation}
where $\vec{P}^\bot$ is the transverse component of $\vec{P}$. The
transverse electrical field $ \vec{E}^\bot $, magnetic induction
$\vec{B}$ and magnetic field $\vec{H} $ are defined by

\begin{equation}\label{d19}
 \vec{E}^\bot=-\frac{\partial\vec{A}}{\partial t},\hspace{1.00
 cm}\vec{B}=\nabla\times\vec{A},\hspace{1.00
 cm}\vec{H}=\frac{\vec{B}}{\mu_0}-\vec{M}.
 \end{equation}

 Using these recent relations, (\ref{d17}) and (\ref{d18}) can be rewritten as
\begin{equation} \label{d20}
\vec{D}=\varepsilon_0 \vec{E}^\bot+\vec{P}^\bot,
\end{equation}
\begin{equation}\label{d21}
\frac{\partial \vec{D}}{\partial t}=\nabla\times\vec{H},
\end{equation}
which are the definitions of the displacement field and the
macroscopic Maxwell equation, in the presence of a
magnetodielectric, respectively.

\subsection{Constitutive equations of the medium}
A magnetodielectric  subjected to electromagnetic field can be
polarized  and magnetized in  consequence of interaction of the
medium with the field. The macroscopic electric and magnetic
polarizations is related to electric and magnetic fields,
respectively by the constitutive equations of the medium. Therefore
a quantization scheme  must be able to give the constitutive
equations in the Heisenberg picture. In this section by applying the
Heisenberg equations to the ladder operators of the medium we find
the correct constitutive equations of the medium.

 the The time evolution of the operators $ d_{\vec{n}\nu}(\vec{q},t) $ and $
b_{\vec{n}\nu}(\vec{q},t)$ can be obtained from the commutation
relations (\ref{d12}) and the Hamiltonian (\ref{d16}) as follows
\begin{eqnarray}\label{d22}
&&\dot{d}_{\vec{n}\nu}(\vec{q},t)=
\frac{\imath}{\hbar}[\tilde{H},d_{\vec{n}\nu}(\vec{q},t)]=\nonumber\\
&&-\imath\omega_{\vec{q}}d_{\vec{n}\nu}(\vec{q},t)+\frac{\imath}{\hbar
\sqrt{V}}\int_V
d^3\vec{r'}e^{-\imath\vec{k}_{\vec{n}}\cdot\vec{r'}}f_{ij}(\omega_{\vec{q}},\vec{r'})
E^i(\vec{r'},t)v^j_{\vec{n}\nu},
\end{eqnarray}

\begin{eqnarray}\label{d23}
&&\dot{b}_{\vec{n}\nu}(\vec{q},t)=
\frac{\imath}{\hbar}[\tilde{H},b_{\vec{n}\nu}(\vec{q},t)]=\nonumber\\
&&-\imath\omega_{\vec{q}}b_{\vec{n}\nu}(\vec{q},t)+\frac{1}{\hbar
\sqrt{V}}\int_V
d^3\vec{r'}e^{-\imath\vec{k}_{\vec{n}}\cdot\vec{r'}}g_{ij}(\omega_{\vec{q}},\vec{r'})
B^i(\vec{r'},t)s^j_{\vec{n}\nu}.
\end{eqnarray}

It is easy to show that these equations have the following formal
solutions

\begin{eqnarray}\label{d24}
&&{d}_{\vec{n}\nu}(\vec{q},t)=
d_{\vec{n}\nu}(\vec{q},0)e^{-\imath\omega_{\vec{q}}t}+\nonumber\\
&&\frac{\imath}{\hbar\sqrt{V}}\int_0^t
dt'e^{-\imath\omega_{\vec{q}}(t-t')} \int_V
d^3r'e^{-\imath\vec{k}_{\vec{n}}\cdot\vec{r'}}f_{ij}(\omega_{\vec{q}},\vec{r'})
E^i(\vec{r'},t')v^j_{\vec{n}\nu},
\end{eqnarray}

\begin{eqnarray}\label{d25}
&&{b}_{\vec{n}\nu}(\vec{q},t)=
b_{\vec{n}\nu}(\vec{q},0)e^{-\imath\omega_{\vec{q}}t}+\nonumber\\
&&\frac{1}{\hbar\sqrt{V}}\int_0^t
dt'e^{-\imath\omega_{\vec{q}}(t-t')} \int_V
d^3r'e^{-\imath\vec{k}_{\vec{n}}\cdot\vec{r'}}g_{ij}(\omega_{\vec{q}},\vec{r'})
B^i(\vec{r'},t')s^j_{\vec{n}\nu}.
\end{eqnarray}

 Substituting (\ref{d24}) in (\ref{d13}) and (\ref{d25}) in (\ref{d14}) we obtain
  the macroscopic constitutive equations of the anisotropic polarizable and magnetizable medium,
\begin{equation}\label{d26}
\vec{P}(\vec{r},t)=\vec{P}_N(\vec{r},t)+\varepsilon_0\int_0^{|t|}
d t' \chi^e(\vec{r},|t|-t')\vec{E}(\vec{r},\pm t'),
\end{equation}
\begin{equation}\label{d27}
\vec{M}(\vec{r},t)=\vec{M}_N(\vec{r},t)+\frac{1}{\mu_0}\int_0^{|t|}
d t' \chi^m(\vec{r},|t|-t')\vec{B}(\vec{r},\pm t'),
\end{equation}
where the upper (lower) sign corresponds to $t>0$ ($ t<0 $) and
$\vec{E}=-\frac{\partial\vec{A}}{\partial
t}-\frac{\vec{P}^\|}{\varepsilon_0} $ is the total electrical field.

The memory tensors
\begin{equation}\label{d28}
 \chi^e(\vec{r},t) =\left\{\begin{array}{cc}
 \frac{8\pi}{\hbar c^3 \varepsilon_0}\int_0^\infty
d\omega\omega^2(ff^t)(\omega,\vec{r})\sin\omega t & \hspace{1.00cm}t>0, \\
\\
0&  \hspace{1.00cm}t\leq 0,  \
  \end{array}\right.
\end{equation}
\begin{equation}\label{d29}
 \chi^m(\vec{r},t) =\left\{\begin{array}{cc}
 \frac{8\pi\mu_0}{\hbar c^3 }\int_0^\infty
d\omega\omega^2(gg^t)(\omega,\vec{r})\sin\omega t & \hspace{1.00cm}t>0, \\
\\
0&  \hspace{1.00cm}t\leq 0,  \
  \end{array}\right.
\end{equation}

are called the electric and magnetic susceptibility tensors
 of the magnetodielectric, respectively and $f^t$, $g^t$ denote the
transpose of the coupling tensors $f$, $g$. If we are given a
definite pair of tensors $\chi^e(\vec{r},t)$, $\chi^m(\vec{r},t)$
which are zero for $t\leq 0 $, then we can inverse (\ref{d28}) and
(\ref{d29}) and obtain the corresponding tensors $(ff^t)$ and
$(gg^t)$ as,

\begin{eqnarray}\label{d30}
&&(ff^t)(\omega,\vec{r})=\nonumber\\
&&\nonumber\\
&&\left\{\begin{array}{cc}
 \frac{\hbar c^3 \varepsilon_0 }{4\pi^2\omega^2}\int_0^\infty
dt\chi^e(\vec{r},t) \sin\omega t=\frac{\hbar c^3 \varepsilon_0
}{4\pi^2\omega^2}Im\left[\underline{\chi}^e(\vec{r},\omega)\right] & \hspace{1.00cm}
\omega>0,\\
\\
  0 & \hspace{1.00cm}\omega=0,
\end{array}\right.\nonumber\\
&&\nonumber\\
&&
\end{eqnarray}
\begin{eqnarray}\label{d31}
&&(gg^t)(\omega,\vec{r})=\nonumber\\
&&\nonumber\\
&&\left\{\begin{array}{cc}
 \frac{\hbar c^3  }{4\pi^2\mu_0\omega^2}\int_0^\infty
dt\chi^m(\vec{r},t) \sin\omega t=\frac{\hbar c^3
}{4\pi^2\mu_0\omega^2}Im\left[\underline{\chi}^m(\vec{r},\omega)\right] &
\hspace{1.00cm} \omega>0,\\
\\
  0 & \hspace{1.00cm}\omega=0,
\end{array}\right.\nonumber\\
&&\nonumber\\
&&
\end{eqnarray}
where $\underline{\chi}^e(\vec{r},\omega)$ and
$\underline{\chi}^m(\vec{r},\omega) $ are the susceptibility tensors
in the frequency domain. The operators $\vec{P}_N $ and $\vec{M}_N$
in (\ref{d26}) and (\ref{d27}) are the noise electric and magnetic
polarization densities

\begin{eqnarray}\label{d32}
&&P_{Ni}(\vec{r},t)=
\frac{1}{\sqrt{V}}\sum_{\vec{n}}\sum_{\nu=1}^3\int
d^3qf_{ij}(\omega_{\vec{q}},\vec{r})\left[d_{\vec{n}\nu}(\vec{q},0)
e^{-\imath\omega_{\vec{q}}t+\imath\vec{k}_{\vec{n}}\cdot\vec{r}}+h.c.\right]
v^j_{\vec{n}
\nu},\nonumber\\
&&
\end{eqnarray}

\begin{eqnarray}\label{d33}
&&M_{Ni}(\vec{r},t)=
\frac{\imath}{\sqrt{V}}\sum_{\vec{n}}\sum_{\nu=1}^3\int
d^3qg_{ij}(\omega_{\vec{q}},\vec{r})\left[b_{\vec{n}\nu}(\vec{q},0)
e^{-\imath\omega_{\vec{q}}t+\imath\vec{k}_{\vec{n}}\cdot\vec{r}}-h.c.\right]
s^j_{\vec{n}
\nu}.,\nonumber\\
&&
\end{eqnarray}

These noises are necessary for a consistent quantization of the
electromagnetic field in the presence of an absorptive medium.

 From (\ref{d30}) and (\ref{d31}) it is clear that for a given pair
 of the susceptibility tensors $ \chi^e$ and $\chi^m$ there are infinite number of
coupling tensors $ f$ and $g$ satisfying the equations (\ref{d30})
and (\ref{d31}). In fact for a given pair of $ \chi^e$ and
$\chi^m$ if the tensors $f$ and $g$ satisfy equations (\ref{d30})
and (\ref{d31}), then the coupling tensors $fA$ and $gA$, for any
orthogonal matrix $(AA^t=1)$, are also a solution. Certainly this
affect the space-time dependence of the noise polarizations and
therefore the space-time dependence of the electromagnetic field
operators, but all of these choices are equivalent. This means
that the various choices of the coupling tensors $ f$ and $g$
satisfying (\ref{d30}) and (\ref{d31}), for a given pair of the
susceptibilities $\chi^e$ and $\chi^m$, do not affect the
commutation relations between the field operators and hence the
physical observables.  This becomes more clear if we compute the
commutation relations between the components of the Fourier
transform of the noise polarizations

\begin{eqnarray}\label{d34}
&&[\underline{P}_{Ni}(\vec{r},\omega) ,
\underline{P}_{Nj}^\dag(\vec{r'},\omega')]=\frac{\hbar\varepsilon_0}{\pi}Im\left[
\underline{\chi}^e_{ij}(\vec{r},\omega)\right]\delta(\vec{r}-\vec{r'})
\delta(\omega-\omega'),\nonumber\\
&&[\underline{M}_{Ni}(\vec{r},\omega) ,
\underline{M}_{Nj}^\dag(\vec{r'},\omega')]=\frac{\hbar}{\mu_0\pi}Im\left[
\underline{\chi}^m_{ij}(\vec{r},\omega)\right]\delta(\vec{r}-\vec{r'})
\delta(\omega-\omega').\nonumber\\
&&
\end{eqnarray}

These relations are generalization of those in reference \cite{19}.
 For a given pair of  $\chi^e $
and $\chi^m $, various choices of the coupling tensors $f$ and $g$
satisfying the relations
  (\ref{d30}) and (\ref{d31}), do not affect these commutation
  relations and accordingly the commutation relations between the
  electromagnetic field operators. Hence, all of the field
  operators which are obtained by using a definite pair of the susceptibilities
  $\chi^e $ and $\chi^m$, with different coupling tensors, satisfying (\ref{d30})
  and (\ref{d31}), are equivalent.

\section{ Solution of Heisenberg equations}
The Maxwell and constitutive equations of the medium constitute a
set of coupled equations. In this section we solve them in terms of
their initial conditions using the Laplace transformation technique.
For any time-dependent operator $ g(t)$ the forward and backward
Laplace transformation of $ g(t)$ are defined by
\begin{eqnarray}\label{d35}
&&g^f(s)=\int_0^\infty dt g(t)e^{-st},\nonumber\\
&&g^b(s)=\int_0^\infty dt g(-t)e^{-st},
\end{eqnarray}

respectively. Carrying out the forward and backward Laplace
transformation of the Maxwell equation (\ref{d21}) and the
constitutive equations (\ref{d20}), (\ref{d26}) and (\ref{d27}) we
find

\begin{eqnarray}\label{d36}
&&\nabla\times\nabla\times\vec{E}^{f,b}(\vec{r},s)+\mu_0\varepsilon_0
s^2\tilde{\varepsilon}(\vec{r},s)\vec{E}^{f,b}(\vec{r},s)-\nonumber\\
&&\nabla\times\tilde{\chi}^m(\vec{r},s)
\nabla\times\vec{E}^{f,b}(\vec{r},s)= \vec{J}^{f,b}(\vec{r},s),
\end{eqnarray}

where $\tilde{\varepsilon}(\vec{r},s)=1+\tilde{\chi}^e(\vec{r},s)$
and $\tilde{\chi}^m(\vec{r},s)$ are the Laplace transformations of
the electric permeability tensor and magnetic susceptibility tensor
of the medium, respectively and

\begin{eqnarray}\label{d37}
&&\vec{J}^{f,b}(\vec{r},s)=\pm\nabla\times\vec{B}(\vec{r},0)-\mu_0s^2
\vec{P}^{f,b}_N(\vec{r},s)\mp\nonumber\\
&&\mu_0s
\nabla\times\vec{M}^{f,b}_N(\vec{r},s)\mp\nabla\times\tilde{\chi}^m(\vec{r},s)
\vec{B}(\vec{r},0)+ \mu_0 s \vec{D}(\vec{r},0),
\end{eqnarray}

is the forward and backward Laplace transformation of the noise
current where upper(lower) sign corresponds to $
\vec{J}^f(\vec{r},s)$ ($\vec{J}^f(\vec{r},s)$). The wave equation
(\ref{d36}) can be solved using the Green tensor method \cite{19}.
To see the space-time dependence of electric field more
explicitly, let us consider a homogeneous but anisotropic bulk
medium. In this case by expanding $\vec{E}^{f,b}(\vec{r},s)$ and
$\vec{J}^{f,b}(\vec{r},s)$ in terms of plane waves as

\begin{eqnarray}\label{d38}
&&\vec{E}^{f,b}(\vec{r},s)=\frac{1}{\sqrt{V}}\sum_{\vec{n}}\underline{\vec{E}}^{f,b}(\vec{k}_{\vec{n}},s)
e^{\imath\vec{k}_{\vec{n}}\cdot\vec{r}},\nonumber\\
&&\vec{J}^{f,b}(\vec{r},s)=\frac{1}{\sqrt{V}}\sum_{\vec{n}}\underline{\vec{J}}^{f,b}(\vec{k}_{\vec{n}},s)
e^{\imath\vec{k}_{\vec{n}}\cdot\vec{r}},\nonumber\\
\end{eqnarray}
and inserting this expansions in the wave equation (\ref{d36}) we
obtain
\begin{eqnarray}\label{d39}
&&\Lambda(\vec{k}_{\vec{n}},s)\underline{\vec{E}}^{f,b}(\vec{k}_{\vec{n}},s)=
\underline{\vec{J}}^{f,b}(\vec{k}_{\vec{n}},s),\nonumber\\
&&\Lambda_{ij}(\vec{k}_{\vec{n}},s)=-\varepsilon_{i\mu\nu}\varepsilon_{\alpha\beta
j}\left[\delta_{\nu\alpha}-\tilde{\chi}^m_{
\nu\alpha}(s)\right]k^\mu_{\vec{n}}k^\beta_{\vec{n}}+\mu_0\varepsilon_0
s^2\tilde{\varepsilon}_{ij}(s),
\end{eqnarray}
where $\varepsilon_{i\mu\nu}$ is the Levi-Civita symbol. we can
use the expansions (\ref{d1}), (\ref{d7}), (\ref{d32}) and
(\ref{d33}) to obtain the operator
$\underline{\vec{J}}^{f,b}(\vec{k}_{\vec{n}},s)$ in terms of the
ladder operators of the electromagnetic field and the
magnetodielectric  medium. Finally using (\ref{d38}) and
(\ref{d39}) after some elaborated calculations we obtain the
space-time dependence of the electric field in terms of the ladder
operators of the field and medium as
\begin{eqnarray}\label{d40}
&&E_i(\vec{r},t)=\sum_{\vec{n}}\sum_{\lambda=1}^2\sqrt{\frac{\hbar\omega_{\vec{n}}
\mu_0}{2cV}}
\left[
\eta^\pm_{ij}(\vec{k}_{\vec{n}},t)a_{\vec{n}\lambda}(0)e^{\imath\vec{k}_{\vec{n}}
\cdot\vec{r}}+h.c.\right]e^j_{\vec{n}\lambda}\nonumber\\
&&-\frac{\mu_0}{\sqrt{V}}\sum_{\vec{n}}\sum_{\lambda=1}^3\int
d^3q\left[
\xi^{\pm}_{ij}(\omega_{\vec{q}},\vec{k}_{\vec{n}},t)d_{\vec{n}\nu}(\vec{q},0)
e^{\imath\vec{k}_{\vec{n}}\cdot\vec{r}}+h.c.\right]v^j_{\vec{n}\nu}\nonumber\\
&&\pm\frac{\mu_0}{\sqrt{V}}\sum_{\vec{n}}\sum_{\lambda=1}^3\int
d^3q\left[
\zeta^{\pm}_{ij}(\omega_{\vec{q}},\vec{k}_{\vec{n}},t)b_{\vec{n}\nu}(\vec{q},0)
e^{\imath\vec{k}_{\vec{n}}\cdot\vec{r}}+h.c.\right]s^j_{\vec{n}\nu},\nonumber\\
&&
\end{eqnarray}

where the upper (lower) sign corresponds to $t>0$ $(t<0)$ and
$\eta^{+}_{ij}(\vec{k}_{\vec{n}},+t)$,
$\eta^{-}_{ij}(\vec{k}_{\vec{n}},-t)$,
$\xi^{+}_{ij}(\omega_{\vec{q}},\vec{k}_{\vec{n}},+t)$, $
\xi^{-}_{ij}(\omega_{\vec{q}},\vec{k}_{\vec{n}},-t)$, $
\zeta^{+}_{ij}(\omega_{\vec{q}},\vec{k}_{\vec{n}},+t)$ and $
\zeta^{-}_{ij}(\omega_{\vec{q}},\vec{k}_{\vec{n}},-t)$ for $ t>0$
are given by

\begin{eqnarray}\label{d41}
&&\eta^{\pm}_{ij}(\vec{k}_{\vec{n}},\pm t)=L^{-1}\left\{
\Lambda_{il}^{-1}(\vec{k}_{\vec{n}},s)\left[\left( \imath s \pm
\omega_{\vec{n}}\right)\delta_{lj}\pm
\omega_{\vec{n}}\varepsilon_{l\mu\nu}\varepsilon_{\alpha\beta
j}\hat{k}^\mu_{\vec{n}}\hat{k}^\beta_{\vec{n}}\tilde{\chi}^m_{
\nu\alpha}(s)\right]\right\},\nonumber\\
 &&\xi^{\pm}_{ij}(\vec{k}_{\vec{n}},\pm t)=L^{-1}\left\{
\Lambda_{il}^{-1}(\vec{k}_{\vec{n}},s)\frac{s^2}{s\pm\imath\omega_{\vec{q}}}
\right\}f_{lj}(\omega_{\vec{q}}),\nonumber\\
&&\zeta^{\pm}_{ij}(\vec{k}_{\vec{n}},\pm t)=L^{-1}\left\{
\Lambda_{il}^{-1}(\vec{k}_{\vec{n}},s)\frac{s}{s\pm\imath\omega_{\vec{q}}}\right\}
\varepsilon_{l\alpha\beta}k_{\vec{n}}^\alpha g_{\beta j}(\omega_{\vec{q}}),\nonumber\\
&&
\end{eqnarray}
and $L^{-1}\{h(s)\}$ denotes the inverse Laplace transformation of
$ h(s)$ and $ \Lambda^{-1}$ is the inverse  of the matrix
$\Lambda$.\\
\textbf{Example 1:}\\
In the first example we show that in the absence of any  medium
this quantization scheme reduces to the usual quantization in the
vacuum. In free space the electric and magnetic susceptibility
tensors are zero and from (\ref{d30}) and (\ref{d31}) we deduce
that the coupling tensors $ f $ and $g$ are also zero. Therefore
from (\ref{d39}), (\ref{d40}) and (\ref{d41}) one finds

\begin{equation}\label{d42}
\vec{E}(\vec{r},t)=\imath\sum_{\vec{n}}\sum_{\lambda=1}^2
\sqrt{\frac{\hbar\omega_{\vec{n}}}{2\varepsilon_0 V}}\left[
a_{\vec{n}\lambda}(0)e^{-\imath\omega_{\vec{n}}
t+\imath\vec{k}_{\vec{n}}\cdot\vec{r}}-h.c.\right]\vec{e}_{\vec{n}\lambda},
\end{equation}

which is the electric field in the free space. So in this case,
quantization of electromagnetic field is
reduced to the usual quantization in the vacuum as expected.\\
\textbf{Example 2 :}\\
Take the susceptibility tensors $\chi^e$ and $\chi^m$ as follows
\begin{displaymath}\label{d43}
\chi^e(\vec{r},t)=\chi^e_0(\vec{r})\times\left\{\begin{array}{ll}
\frac{1}{\triangle} & 0<t<\triangle,\\
\\
0 & \textrm{otherwise},
\end{array}\right.
\end{displaymath}
\begin{displaymath}\label{d44}
\chi^m(\vec{r},t)=\chi^m_0(\vec{r})\times\left\{\begin{array}{ll}
\frac{1}{\triangle} & 0<t<\triangle,\\
\\
0 & \textrm{otherwise},
\end{array} \right.
\end{displaymath}

where $\chi^e_0(\vec{r})$ and $\chi^m_0(\vec{r})$ are some time
independent but position dependent tensors and $\triangle $ is a
real positive constant, using (\ref{d30}) and (\ref{d31}) we find

\begin{eqnarray}\label{d45}
(ff^t)(\omega,\vec{r})=\frac{\hbar c^3\varepsilon_0}{4\pi^2
\omega^2}\frac{\sin^2\frac{\omega\triangle}{2}}
{\frac{\omega\triangle}{2}}\chi^e_0(\vec{r}),\nonumber\\
(gg^t)(\omega,\vec{r})=\frac{\hbar c^3}{4\pi^2
\omega^2\mu_0}\frac{\sin^2\frac{\omega\triangle}{2}}
{\frac{\omega\triangle}{2}}\chi^m_0(\vec{r}),\nonumber\\
\end{eqnarray}

and from (\ref{d26}) and (\ref{d27})

\begin{eqnarray}\label{d46}
&&\vec{P}(\vec{r},t)=\vec{P}_N(\vec{r},t)+\chi^e_0(\vec{r})\frac{\varepsilon_0}
{\triangle}
\int_{|t|-\triangle}^{|t|} d t'\vec{E}(\vec{r},\pm t'),\nonumber\\
&&\vec{M}(\vec{r},t)=\vec{M}_N(\vec{r},t)+\chi^m_0(\vec{r})\frac{1}{\mu_0
\triangle}\int_{|t|-\triangle}^{|t|} d t'\vec{B}(\vec{r},\pm t'),
\end{eqnarray}

where $\vec{P}_N(\vec{r},t),\vec{M}_N(\vec{r},t)$ are the noise
polarization densities correspond to a pair coupling tensors $f$ and
$g$ satisfying (\ref{d45}). In the limit $ \triangle \rightarrow 0
$, from (\ref{d45}) we deduce that the coupling tensors and
therefore the noise polarization densities defined by (\ref{d32})
and (\ref{d33}) tend to zero. In this case the constitutive
equations (\ref{d46}) are

\begin{eqnarray}\label{d47}
\vec{P}(\vec{r},t)&=&\varepsilon_0\chi^e_0(\vec{r})\vec{E}(\vec{r},t),\nonumber\\
\vec{M}(\vec{r},t)&=&\frac{1}{\mu_0}\chi^m_0(\vec{r})\vec{B}(\vec{r},t),\nonumber\\
\end{eqnarray}

and the wave equation (\ref{d36}) becomes

\begin{eqnarray}\label{d48}
&&\nabla\times\nabla\times\vec{E}^{f,b}(\vec{r},s)+\mu_0\varepsilon_0
s^2\left[1+\chi^e_0(\vec{r})\right]\vec{E}^{f,b}(\vec{r},s)-\nonumber\\
&&\nabla\times\chi^m_0(\vec{r})
\nabla\times\vec{E}^{f,b}(\vec{r},s)= \vec{J}^{f,b}(\vec{r},s),
\end{eqnarray}

where the noise current density (\ref{d37}) is

\begin{eqnarray}\label{d49}
&&\vec{J}^{f,b}(\vec{r},s)=\pm\nabla\times\vec{B}(\vec{r},0)\mp
\nabla\times\chi^m_0(\vec{r})\vec{B}(\vec{r},0)+ \mu_0 s\vec{D}(\vec{r},0).\nonumber\\
&&
\end{eqnarray}

We see that the noise operators have vanished. This is because in
the limit $\triangle\rightarrow 0$, the absorption coefficients
tend to zero. The solution of the wave equation (\ref{d48}) can be
expressed in terms of the Green tensor as

\begin{equation}\label{d50}
\vec{E}^{f,b}(\vec{r},s)=\int
d^3r'G(\vec{r},\vec{r'},s)\vec{J}^{f,b}(\vec{r'},s),
\end{equation}

where the Green tensor satisfies the equation
\begin{eqnarray}\label{d51}
&&\nabla\times\nabla\times
G(\vec{r},\vec{r'},s)+\mu_0\varepsilon_0
s^2\left[1+\chi^e_0(\vec{r})\right]G(\vec{r},\vec{r'},s)-\nonumber\\
&&\nabla\times\chi^m_0(\vec{r}) \nabla\times
G(\vec{r},\vec{r'},s)=\delta(\vec{r}-\vec{r'}),
\end{eqnarray}

together with some boundary conditions. These boundary conditions
 guarantee the continuity of the tangential component of
electric field and the normal component of magnetic field at some
surfaces where the susceptibilities of the medium become
discontinuous. For an anisotropic homogeneous medium using
(\ref{d39}), (\ref{d40}) and (\ref{d41}) we can write the electric
field as follows

\begin{eqnarray}\label{d51.1}
&&E_i(\vec{r},t)=\sum_{\vec{n}}\sum_{\lambda=1}^2\sqrt{\frac{\hbar\omega_{\vec{n}}
\mu_0}{2cV}}
\left[
\eta^\pm_{ij}(\vec{k}_{\vec{n}},t)a_{\vec{n}\lambda}(0)e^{\imath\vec{k}_{\vec{n}}
\cdot\vec{r}}+h.c.\right]e^j_{\vec{n}\lambda},\nonumber\\
\end{eqnarray}

where

\begin{eqnarray}\label{d51.2}
&&\eta^{\pm}_{ij}(\vec{k}_{\vec{n}},\pm t)=L^{-1}\left\{
\Lambda_{il}^{-1}(\vec{k}_{\vec{n}},s)\left[\left( \imath s \pm
\omega_{\vec{n}}\right)\delta_{lj}\pm
\omega_{\vec{n}}\varepsilon_{l\mu\nu}\varepsilon_{\alpha\beta
j}\hat{k}^\mu_{\vec{n}}\hat{k}^\beta_{\vec{n}}(\chi^m_0)_{
\nu\alpha}\right]\right\},\nonumber\\
&&\Lambda_{il}(\vec{k}_{\vec{n}},s)=-\varepsilon_{i\mu\nu}\varepsilon_{\alpha\beta
l}\left[\delta_{\nu\alpha}-(\chi^m_0)_{
\nu\alpha}\right]k^\mu_{\vec{n}}k^\beta_{\vec{n}}+\mu_0\varepsilon_0
s^2\left(1+(\chi^e_0)_{il}\right).
\end{eqnarray}

 This example shows that the present quantization
scheme is reduced to the usual quantization in a nonabsorptive
medium.\\
\textbf{Example 4: A simple model for electric susceptibility tensor}\\
If we neglect the difference between local and macroscopic electric
field for substances with a low density, then the classical equation
of a bound atomic electron in an external electric field for small
oscillation can be written as

\begin{equation}\label{d52}
\ddot{x}_i+2\gamma\dot{x}_i+K_{ij}x_j=-\frac{e}{m}\vec{E}_i(t),\hspace{2.00
cm}i=1,2,3,
\end{equation}

where $\vec{E}(t)$ is the electric field in the place of the atom
and the magnetic force has been neglected in comparison with the
electric one and $\gamma$ is a damping coefficient. We have assumed
that for sufficiently small oscillations the ith component of the
force exerted on the bound electron by nucleus, can be expressed as
a linear combination of the coordinates of the electron with
constant coefficients $K_{ij}$. Therefore in this simple model the
motion of the bound electron is as a forced anisotropic harmonic
oscillator. Let $\underline{\vec{E}}(\omega) $ and
$\underline{\vec{r}}(\omega)$ be Fourier transforms of the
electrical field $\vec{E}(t)$ and position $\vec{r}(t) $
respectively. From (\ref{d52}) we  find

\begin{equation}\label{d53}
\underline{\vec{r}}(\omega)=
=-\frac{e}{m}\left[(-\omega^2+2\imath\gamma
\omega)1+K\right]^{-1}\underline{\vec{E}}(\omega).
\end{equation}

Now let there be $ N $ molecules per unit volume of the medium with
$ z $ electrons per molecule. We assume that the damping coefficient
($\gamma$) and the tensor $K$ are identical for each electron. Then
for the Fourier transform of the electric polarization density we
find

\begin{equation}\label{d54}
\underline{\vec{P}}(\vec{r},\omega)=
\frac{Ne^2}{m}\left[(-\omega^2+2\imath\gamma
\omega)1+K\right]^{-1}\underline{\vec{E}}(\vec{r},\omega).
\end{equation}

From (\ref{d54}) we find the electric susceptibility tensor of the
medium in frequency domain

\begin{equation}\label{d55}
\underline{\chi}^e(\omega)=\frac{Ne^2}{m\varepsilon_0}\left[(-\omega^2+2\imath\gamma
\omega)1+K\right]^{-1}.
\end{equation}

 From (\ref{d30}) we have

 \begin{eqnarray}\label{d56}
 (ff^t)(\omega)=\left\{\begin{array}{cc}
   \frac{\hbar c^3\varepsilon_0}{4\pi^2\omega^2}2\gamma\omega\left[
   \left(-\omega^21+K\right)^2+4\gamma^2\omega^21\right]^{-1} & \omega\neq0, \\
   \\
   0 & \omega=0 \
 \end{array}\right.
 \end{eqnarray}

In the special case $\gamma=0 $, the anisotropic dielectric
substance is a nonabsorptive  one and this relation for
$\omega\neq0$ becomes
\begin{equation}\label{d57}
(ff^t)(\omega)=\frac{\hbar
c^3\varepsilon_0}{4\pi\omega^2}\sum_{i=1}^3\delta(\omega-\omega_i)
\left(\begin{array}{ccc}
  x_i^2 & x_iy_i & x_iz_i \\
  y_ix_i & y_i^2 & y_iz_i \\
  z_ix_i & z_iy_i & z_i^2
\end{array}\right).
\end{equation}

where $\omega_i$, are eigenvalues of the tensor $K$ corresponding
to eigenvectors $ R_i= \left(\begin{array}{c}
  x_i \\
  y_i \\
  z_i
\end{array}\right)$,$\hspace{0.3cm}(i=1,2,3)$. In this
case the coupling tensor $f$ is nonzero only for frequencies
$\omega_1,\omega_2,\omega_3 $,. These frequencies are
 the resonance frequencies  of the equation (\ref{d52}). Therefore
  when $\gamma=0$, that is for a non absorptive
medium, the coupling tensor $f$ and therefore the noise electrical
polarization density is equal to zero except in resonance
frequencies, where the energy of electromagnetic field is absorbed
by the oscillator. This example explicitly shows that this
quantization scheme is also applicable to anisotropic dispersive
but non absorptive media.
\section{Summary}
The electromagnetic field quantization in the presence of an
anisotropic magnetodielectric is investigated consistently by
modelling the magnetodielectric with two independent quantum
fields namely E and M quantum fields. For a given pair of the
electric and magnetic susceptibility tensors $\chi^e$ and
$\chi^m$, we have found the corresponding coupling tensors $f$ and
$g$, which couple electromagnetic field to E and M quantum fields
respectively. The explicit space-time dependence of the noise
polarizations are obtained in terms of the ladder operators of the
medium and the coupling tensors as a consequence of Heisenberg
equations. Maxwell and constitutive equations are obtained
directly from Heisenberg equations. In the limiting case, i.e.,
when there is no medium, the approach tends to the usual method of
quantization of the electromagnetic field in vacuum. Also when the
medium is a non absorptive one, the noise polarizations tend to
zero and in this case the approach is equivalent to the previous
methods, as expected.

\end{document}